\documentclass[journal]{IEEEtran}
\usepackage{glossaries}

\usepackage{graphicx} 
\usepackage{stfloats}
\usepackage{enumerate}%
\usepackage{graphicx} 
\usepackage{enumerate}%
\usepackage{bm}
\usepackage{multicol}
\usepackage[export]{adjustbox}
\usepackage{float}
\usepackage{graphicx} 
\usepackage{caption}
\usepackage[labelformat=simple]{subcaption}

\usepackage[font={footnotesize}]{caption, subcaption}
\usepackage[ruled,vlined]{algorithm2e}

\usepackage{caption}
\usepackage[labelformat=simple]{subcaption}

\usepackage[font={footnotesize}]{caption, subcaption}
\PassOptionsToPackage{bookmarks={false}}{hyperref}

\usepackage{amssymb}

\usepackage{bm}

\usepackage{cite}

\ifCLASSINFOpdf
\else
\fi
\begin{document}

\title{Adaptive Turbo Equalization for Nonlinearity Compensation in WDM Systems}
%
%
%

\author{Edson~P.~da~Silva,~\IEEEmembership{Senior Member,~IEEE,~OSA}, Metodi P. Yankov,~\IEEEmembership{Member,~IEEE}

\thanks{Edson~P.~da~Silva is with the Department of Electrical Engineering, Federal University of Campina Grande (UFCG), Campina Grande,
Paraíba, Brazil, e-mails: edson.silva@dee.ufcg.edu.br. \\ Metodi P. Yankov is with the Department of Photonics Engineering (Fotonik) of the Technical University of Denmark (DTU), 2800 Kgs. Lyngby, Denmark.}

\thanks{Manuscript received April, 2021; revised .}\vspace{-0.5cm}}

%
%

\markboth{Journal of Lightwave Technology, Class Files,~Vol.~X, No.~X, April~2021}%
{Shell \MakeLowercase{\textit{et al.}}: Bare Demo of IEEEtran.cls for Journals}
%



\maketitle

\bstctlcite{IEEEexample:BSTcontrol}

\begin{abstract}
In this paper, the performance of adaptive turbo equalization for nonlinearity compensation (NLC) is investigated. A turbo equalization scheme is proposed where a recursive least-squares (RLS) algorithm is used as an adaptive channel estimator to track the time-varying intersymbol interference (ISI) coefficients associated with inter-channel nonlinear interference (NLI) model. The estimated channel coefficients are used by a MIMO $2\times2$ soft-input soft-output (SISO) linear minimum mean square error (LMMSE) equalizer to compensate for the time-varying ISI. The SISO LMMSE equalizer and the SISO forward error correction (FEC) decoder exchange extrinsic information in every turbo iteration, allowing the receiver to improve the performance of the channel estimation and the equalization, achieving lower bit-error-rate (BER) values. The proposed scheme is investigated for polarization multiplexed 64QAM and 256QAM, although it applies to any proper modulation format. Extensive numerical results are presented. It is shown that the scheme allows up to 0.7~dB extra gain in effectively received signal-to-noise ratio (SNR) and up to 0.2~bits/symbol/pol in generalized mutual information (GMI), on top of the gain provided by single-channel digital backpropagation.
\end{abstract}

\begin{IEEEkeywords}
Nonlinearity Compensation, Turbo Equalization, Digital Backpropagation.
\end{IEEEkeywords}

\IEEEpeerreviewmaketitle

\section{Introduction}\label{secI}
\IEEEPARstart{T}HE interplay of Kerr nonlinearity, chromatic dispersion (CD), and amplified spontaneous emission (ASE) causes signal distortions that accumulate as the optical carriers propagate through the optical fiber channel. The overall result of such phenomena is popularly known in the literature as nonlinear interference (NLI) \cite{Dar2013}. For fiber channels with large accumulated and uncompensated dispersion \cite{Poggiolini2014}, a standard coherent optical receiver perceives the NLI as a Gaussian noise source. The presence of NLI reduces the effectively received signal-to-noise ratio (SNR) at the input of the receiver, hence penalizing the transmission performance. The variance of the NLI is directly proportional to the cube of signal power launched into the fiber, and it is responsible for the decreasing of the achievable information rate (AIR)\cite{Alvarado2018} over the fiber channel when the power sent into the fiber is set above a certain threshold, known as the optimal launch power \cite{Poggiolini2014}. For a given fiber channel, the AIR curve as a function of the fiber input power is popularly known as \textit{the nonlinear Shannon limit} \cite{Essiambre2010}. These curves represent lower bounds on the capacity of the fiber channel, which remains unknown to this date. The NLI is recognized as the dominating impairment limiting the AIRs over standard single-mode fiber (SMF) channels.

The need for increased throughput and spectral-efficiency in wavelength division multiplexing (WDM) systems has pushed the research to look for solutions to overcome the limits set by Kerr nonlinearity. To this end, the study of nonlinearity compensation (NLC) methods based on digital signal processing (DSP) is one of the most explored research topics of the last decade \cite{Cartledge2017} in the optical communications community. However, practical and effective implementation of NLC based on DSP is challenging. Particularly, due to the large computational complexity of the NLC algorithms, which reflects the complicated characteristics of NLI and the long memory induced by CD in the fiber channel. 

Considering the nonlinear processes that generate the NLI, those distortions can be classified into two categories: deterministic and stochastic \cite{Dar2015}. The deterministic NLI is produced by interactions involving the propagating signals and the nonlinear medium. In principle, these distortions can be computed, if all the necessary information about the interacting signals and the physical parameters of the fiber channel is known. The stochastic NLI is produced when the same processes that generate the deterministic distortions are affected by the presence of ASE or random channel fluctuations. Hence, the stochastic NLI produces distortions with random and time-varying nature, restricting the strategies that the receiver can adopt to compensate for it. The majority of the NLC methods aim for compensating the deterministic fraction of the distortions via zero-forcing equalization \cite{Ghazisaeidi2019}, while only the slow, time-varying component of the stochastic NLI can be targeted by adaptive equalization \cite{Dar2015}. The stochastic NLI is considered to be the ultimate impairment limiting the rates over the optical fiber channel.

Considering the receiver's capabilities in detecting and processing the optical fields, it is also common to categorize the NLI into intra-channel or inter-channel. The intra-channel NLI is generated by the mixing of frequency components within the bandwidth of the detected WDM channel. The inter-channel NLI is produced by the mixing of frequency components from signals at different WDM channels. From the receiver side, the inter-channel NLI is either seen as deterministic or stochastic, depending on whether the receiver has access or not to the detected field of all co-propagating WDM signals, which is rarely the case. Moreover, when the number of co-propagating WDM signals increases, the inter-channel effects dominate the NLI generation, making the inter-channel NLI the main impairment limiting the achievable rates in practice. 

Several different algorithms have been proposed to compensate for the NLI \cite{Cartledge2017}. Due to its simplicity and effectiveness, digital backpropagation (DBP)\cite{Roberts2006, Ip2010a, Ip2010, Temprana2015a} is the most assessed algorithm for DSP-based NLC. Acting as a zero-forcing equalizer for the in-band deterministic NLI, DBP relies on the split-step Fourier method (SSFM) to simulate the backward propagation of the signal over the nonlinear fiber channel. Ideally, the backward propagation reverses all deterministic NLI, increasing the SNR available at the receiver and, consequently, the AIRs. The algorithm is flexible enough to implement either intra-channel or inter-channel NLC. However, so far, effective inter-channel NLC with DBP requires a prohibitively large amount of computational resources to allow any practical implementation.

Interestingly, it has been shown that a single-channel coherent receiver will see part of the inter-channel NLI manifesting itself as a time-varying intersymbol interference (ISI) \cite{Secondini2013a, Dar2015}. This finding has unveiled the possibility of using adaptive equalization to make the receiver able to partially compensate for inter-channel NLI without the knowledge of the data carried by the interfering WDM channels. Since then, this problem has been discussed in the literature by a number of authors\cite{Secondini2014a,Golani2018_2, Golani2018, Golani2019, DaSilva2019, Paskov2019}. Particularly, the tracking of the fast time-varying ISI dynamics is challenging, requiring more advanced algorithms compared to the standard adaptive equalizers used in coherent optical receivers.

In this work, we propose and investigate an adaptive turbo equalization scheme to implement inter-channel NLI compensation in single-channel dual-polarization coherent optical receivers. The proposed scheme relies on turbo iterations between a soft-input soft-output (SISO) equalizer assisted by an adaptive recursive least squares (RLS) channel estimator and a SISO low-density parity-check (LDPC) decoder. In the following, we discuss related work and list the original contributions of this work.

\subsection{Related work}\label{subSecI.a}

Similar to the approach followed in this work, other turbo equalization algorithms have been proposed before. 

In \cite{Arlunno2014}, turbo equalization is considered as a general tool for impairment mitigation in coherent optical receivers. However, the authors do not investigate any scheme specifically tailored to compensate for inter-channel NLI. Moreover, only convolutional codes are considered.

In \cite{Pan2015}, a code-aided turbo carrier phase recovery (CPR) algorithm is proposed and experimentally investigated for 16QAM signals. The authors demonstrate performance gains by improved compensation of the nonlinear phase noise. The proposed algorithm iterates between CPR and FEC decoding, using soft-information provided by the LDPC decoder to enhance the performance of the CPR.

In \cite{Golani2018}, the authors investigate potential gains to be achieved with the use of adaptive equalization for inter-channel NLI compensation. The authors also propose an equalization scheme based on Kalman filtering and maximum likelihood sequence estimation (MLSE) to achieve improved performances. However, such a receiver scheme is proposed more like an idealized model to estimate bounds on performance improvement, rather than considering its integration with more standard processing used in coherent receivers. Due to the large complexity of the model, the authors limit their analysis to 16QAM signals.

In \cite{Golani2019}, the authors propose a turbo equalization scheme for inter-channel NLI compensation. For that, a Kalman algorithm to perform channel estimation is used, providing the time-varying channel coefficients to a linear minimum mean square error (LMMSE) equalizer. Soft information is feedback from the LDPC decoder to the channel estimator such that, at every iteration in the turbo loop, better information of the channel state is provided to the LMMSE equalizer, iteratively improving the overall equalization performance. The authors report improvements of up to $\mathrm{1.3~dB}$ in effectively received SNR and up to $\mathrm{0.3~bits/symbol/pol}$. in AIR for 64QAM. In \cite{Golani2019_2}, the same authors further extend their analysis comparing different equalization approaches showing that better performances are achieved by their turbo equalization with Kalman-based channel estimation. However, in \cite{Golani2019,Golani2019_2}, an ideal DSP chain is assumed, such that the influence of the standard adaptive equalizer and CPR are not taken into account. Moreover, the proposed Kalman channel estimation algorithms are considerably more complex than e.g. the RLS equalizer algorithms the authors use as benchmark for performance comparison.

In \cite{DaSilva2019}, the perturbation-based intra-channel NLC is combined with an RLS adaptive equalizer for inter-channel NLC. In the proposed configuration, the receiver iterates between NLC and FEC decoding, with the NLC algorithms being assisted with hard-decision feedback from the forward error correction (FEC) decoder. Simulations and experiments showed that establishing this iterative feedback from the FEC decoder to the equalization algorithms enables improved NLC performance. Results are shown for 16 and 64QAM. However, since the feedback from the decoder to the equalizers is based on hard-decisions, this scheme can only be considered as a very simplified version of a turbo equalizer.

In \cite{KoikeAkino2020}, an NLC turbo equalization scheme based on a neural network is proposed. Different QAM formats are analyzed up to 64QAM. The proposed algorithm is shown to increase  AIRs by up to $\mathrm{0.12~b/s/Hz}$. However, the authors consider only non-standard transmission links with dispersion management and non-zero dispersion-shifted fiber spans, which are known to increase substantially the NLI. Moreover, the proposed deep neural-network-based turbo equalizer increases significantly the complexity of the receiver. The usually long time required to train neural-network also makes it difficult applying such approach for equalization of time-varying channels.

\subsection{Contributions of this work}
The main contributions of this work are summarized as follows:

\begin{enumerate}
\item \textit{RLS adaptive filter as channel estimator}: in all previous references \cite{Dar2015, Golani2018, Golani2019, Golani2019_2, DaSilva2019}, the RLS adaptive filter was used as an adaptive equalizer. The results presented in those references are in agreement, showing performance improvements around $\mathrm{0.2~dB}$ in effectively received SNR by using the RLS adaptive equalizer to compensate for the NLI induced time-varying ISI. Instead, in this work it is proposed to use the RLS adaptive filter as an \textit{adaptive channel estimator}. The adaptive equalization is passed to an LMMSE equalizer. Therefore, similarly to \cite{Golani2019}, here the tasks of adaptive channel estimation and adaptive equalization are split. This modification in the function of the RLS adaptive filter is shown to improve the compensation of the time-varying ISI, allowing up to $\mathrm{0.7~dB}$ gain in effectively received SNR. Moreover, using an RLS adaptive filter instead of the Kalman algortithm proposed in \cite{Golani2019} greatly reduces the complexity of the channel estimation algorithm.

\item \textit{LMMSE equalization with \textit{a priori} information on the received symbols}: in a similar fashion of what \cite{Tuchler2002} proposes, and differently from the scheme proposed in \cite{Golani2019}, in this paper the adaptive LMMSE equalizer taps are calculated considering the available soft-information on received symbols (i.e. means and variances) obtained with the output of the SISO decoder.

\item \textit{Performance analysis considering a standard receiver DSP chain}: in \cite{Golani2019} the authors test their proposed turbo equalization scheme assuming an ideal coherent receiver and, hence, not considering the influence on the NLC of e.g. the adaptive equalizer used for polarization demultiplexing and the CPR block used in a standard receiver DSP chain. However, it is known that those blocks are able to compensate fractions of the time-varying NLI \cite{Pan2015,Yankov2015,Yankov2017,Fehenberger2015}. Consequently, the gains indicated in \cite{Golani2019} could be smaller if the algorithm is implemented after a standard receiver DSP chain. Here, the proposed turbo equalization scheme is implemented after a pilot-based receiver DSP chain that includes a $T_s/2$-fractionally spaced adaptive equalizer and a carrier phase recovery algorithm. Hence, the results reported in this paper already account for the more realistic scenario where part of the time-varying NLI could be eliminated before the received signal reaches the turbo equalizer.

\item \textit{Transmission reach}: the transmission reach extension enabled by the proposed scheme is investigated. When compared with single-channel DBP, the results show that turbo equalization is capable of providing similar additional benefits in terms of reach extension.
\end{enumerate}

The remaining of the paper is divided as follows: in Section~\ref{SecII}, a detailed description of the proposed turbo equalization scheme is given; in Section~\ref{SecIII}, the performance of the proposed method is investigated via numerical simulations and the results are discussed, leading to the conclusions.

\subsubsection*{Notation}
In what follows, we use $\operatorname{E}\{.\}$ to denote expectation, $^T$ and $^H$ the transpose and conjugate transpose, respectively. Bold uppercase letters indicate matrices, while bold lowercase letters indicate vectors, except for $\mathrm{2\times1}$ dual polarization vectors. Anytime needed, polarization components are distinguished by superscripts $^{(x)}$ and $^{(y)}$.

\begin{figure*}[b]
\centering
\includegraphics[width=0.9\linewidth]{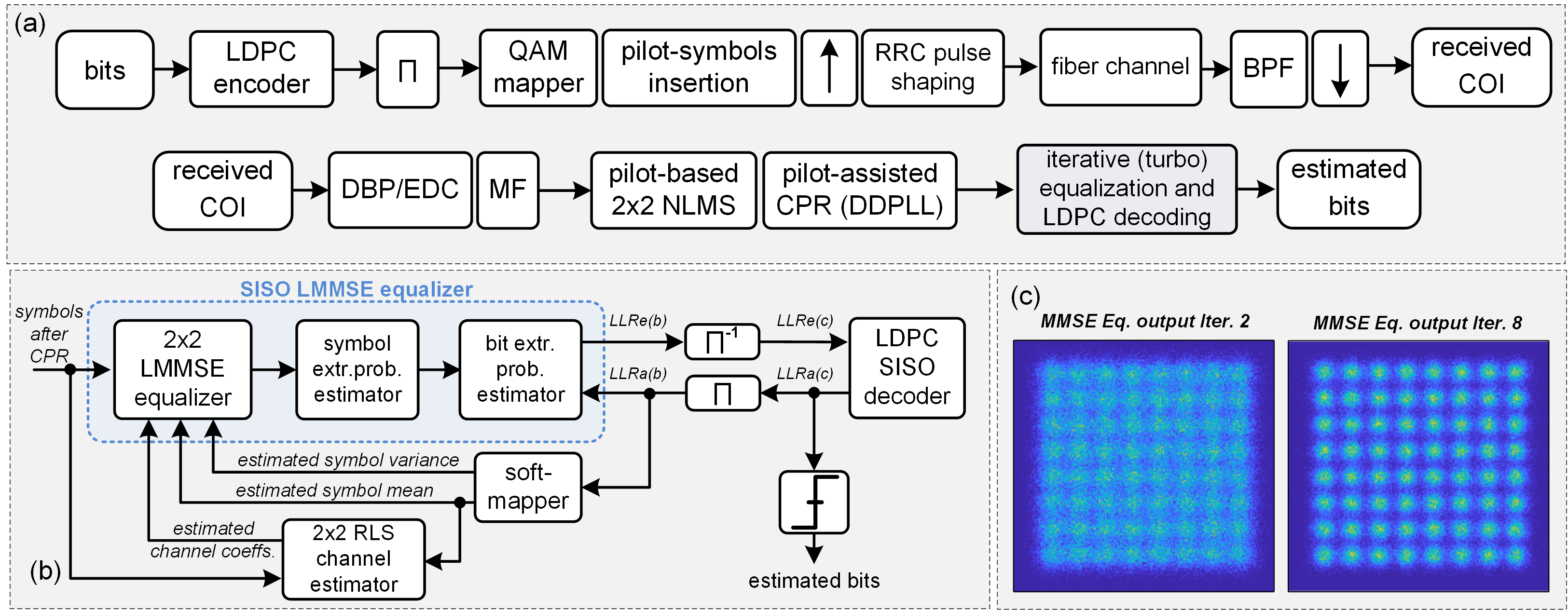}
\caption{(a) DSP chain at the transmitter and receiver. RRC: root-raised cosine, BPF: band-pass filter, COI: channel of interest, MF: matched filter; (b) Detailed block diagram of the turbo equalization scheme; (c) Example of the constellation at the output of the LMMSE equalizer at different iterations showing the improvement in SNR.}
\label{Fig1}
\end{figure*}

\section{Iterative NLC using turbo equalization}\label{SecII}

The transmitter and receiver block diagrams considered in this work are shown in Fig.~\ref{Fig1}~(a). The proposed turbo equalization algorithm is depicted in Fig.~\ref{Fig1}~(b). At each turbo iteration, the SISO equalizer and the SISO LDPC decoder exchange updated soft information on the received bit sequences. The updates from the decoder are used to improve the channel estimation and the LMMSE equalization. The updated extrinsic output of equalizer will help the decoder to achieve lower bit-error-rates (BERs) until the performance saturation of the iterative process. The proposed SISO equalizer is a generalized version of the algorithm proposed in \cite{Dejonghe2002} to a time-varying MIMO $\mathrm{2\times 2}$ channel.

In what follows, the processing steps comprising the SISO equalizer and the RLS adaptive channel estimation shown in Fig.~\ref{Fig1}~(b) are discussed in detail.

\subsection{The SISO equalizer}\label{secII.A}
The SISO equalizer takes as input the received symbols, the time-varying estimated channel coefficients, and the \textit{a priori} soft information available on the received symbol sequence. This input is then used to perform LMMSE equalization and produce the \textit{a posteriori} extrinsic information output which will be forward to the SISO LDPC decoder.
\\
\subsubsection{Estimating first and second order statistics of the symbols}

Assume that $s_{j}$ is a transmitted symbol from a given \textit{proper} QAM constellation \cite{Taubock2012b}, i.e. $s_{j} \in \mathcal{S}$, where $\mathcal{S}$ is the set of all $M$ constellation symbols. Assuming independence between the interleaved coded bits, the \textit{a priori} probability $P_{a}\left(s_{j}\right)$ of $s_{j}$ can be written in terms of \textit{a priori} bit probabilities $P_{a}\left(b_{j}^{l}\right)$ as 
\begin{equation}
P_{a}\left(s_{j}\right)=\prod_{l=1, \ldots, q} P_{a}\left(b_{j}^{l}\right),
\end{equation}
\noindent where $q=\log_2M$ and $\left\lbrace b_{j}^{l} \right\rbrace_{l=1}^{q}$ is the bit sequence mapped to $s_{j}$.

The bit \textit{a priori} probabilities can be calculated from the bit \textit{a priori} \textit{L-values} \cite{Alvarado2015}, defined as
\begin{equation}
L_{a}\left(b_{j}^{l}\right)=\ln \frac{P_{a}\left(b_{j}^{l}=1\right)}{P_{a}\left(b_{j}^{l}=0\right)}.
\end{equation}
With the \textit{a priori} information available at the instant $j$, one can calculate the first and second order statistics of the symbols. Thus, the mean $\bar{s}_{j}$ and variance $\operatorname{var}\{s_{j}\}$ of $s_{j}$ can be obtained as 
\begin{equation}
\bar{s}_{j} \triangleq \operatorname{E}\left\{s_{j}\right\}=\sum_{s_{j} \in \mathcal{S}} s_{j} P_{a}\left(s_{j}\right),
\end{equation}
\begin{equation}
\operatorname{var}\left\{s_{j}\right\} \triangleq \operatorname{E}\left\{\left|s_{j}\right|^{2}\right\}-\left|\bar{s}_{j}\right|^{2}, 
\end{equation}
\noindent where 
\begin{equation}
\operatorname{E}\left\{\left|s_{j}\right|^{2}\right\}=\sum_{s_{j} \in \mathcal{S}}\left|s_{j}\right|^{2} P_{a}\left(s_{j}\right).
\end{equation}

\noindent Here, the QAM constellations are assumed to be square with unitary average energy per symbol such that, without prior information on $s_j$, we have $\operatorname{E}\left\{s_{j}\right\}=0$ and $\sigma_s^2~=~\operatorname{var}\left\{s_{j}\right\}~=~1$.
\\
\subsubsection{MIMO LMMSE equalizer with \textit{a priori} soft information}

Define the equalizer length as $N \triangleq N_{1}+N_{2}+1$. The channel model accounting for the time-varying ISI induced by inter-channel NLI is represented by the MIMO $2\times2$ sliding-window model where, at each time step $i$,  the received dual-pol. signal is given by
\begin{equation}
\mathbf{r}_{i}=\mathbf{H}_i \mathbf{s}_{i}+\mathbf{n}_{i},
\end{equation}

\noindent with the vectors of received symbols, transmitted symbols and noise represented by $\mathbf{r}_{i}~=~\left[\mathbf{r}_{i}^{(x)}, \mathbf{r}_{i}^{(y)} \right]^T$, $\mathbf{s}_{i}~=~\left[\mathbf{s}_{i}^{(x)}, \mathbf{s}_{i}^{(y)} \right]^T$, and $\mathbf{n}_{i}~=~\left[\mathbf{n}_{i}^{(x)}, \mathbf{n}_{i}^{(y)} \right]^T$, respectively. The superscripts $p\in \{x,y\}$ indicate the polarization components, and  
\begin{equation}
\begin{aligned}
\mathbf{r}_{i}^{(p)} & \triangleq\left[\begin{array}{lllll}
r_{i-N_{1}}^{(p)}, & \ldots, & r_{i}^{(p)}, & \ldots, & r_{i+N_{2}}^{(p)}
\end{array}\right]_{N \times 1}, \\
\mathbf{s}_{i}^{(p)} & \triangleq\left[\begin{array}{lllll}
s_{i-N_{1}-L}^{(p)}, & \ldots, & s_{i}^{(p)}, & \ldots, & s_{i+N_{2}}^{(p)}
\end{array}\right]_{(N+L) \times 1}, \\
\mathbf{n}_{i}^{(p)} & \triangleq\left[\begin{array}{lllll}
n_{i-N_{1}}^{(p)}, & \ldots, & n_{i}^{(p)}, & \ldots, & n_{i+N_{2}}^{(p)}
\end{array}\right]_{N \times 1}.
\end{aligned}
\end{equation}

The time-varying channel matrix $\mathbf{H}_i$ has size $(2N \times(2N+2L))$ and it is given by
\begin{equation}
\mathbf{H}_i \triangleq\left[\begin{array}{cc}
\mathbf{H}_i^{(xx)} & \mathbf{H}_i^{(xy)} \\
\mathbf{H}_i^{(yx)} & \mathbf{H}_i^{(yy)} \\
\end{array}\right],
\end{equation}
\noindent where each sub-matrix has size $(N \times(N+L))$ and it is given by
\begin{equation}
\mathbf{H}_i^{k} \triangleq\left[\begin{array}{ccccccc}
h_{L}^{(k)} & \ldots & h_{0}^{(k)} & 0 & \ldots & \ldots & 0 \\
0 & h_{L}^{(k)} & \ldots & h_{0}^{(k)} & 0 & \ldots & 0 \\
\vdots & \ddots & \ddots & \ddots & \ddots & \ddots & \vdots \\
0 & \ldots & \ldots & 0 & h_{L}^{(k)} & \ldots & h_{0}^{(k)}
\end{array}\right],
\end{equation}

\noindent where $\left\lbrace h_{n}^{(k)} \right\rbrace_{n=0}^{L}$ are the estimated channel coefficients at the time step $i$, with $k\in \{xx,\; xy,\; yx,\; yy\}$. Finally, the complex-valued noise samples of each polarization are assumed to be uncorrelated, circular, and Gaussian distributed, i.e. $\mathbf{n}_{i}^{(p)}~\sim~\mathcal{N}_{c}\left(0, \sigma_{n}^{2} \mathbf{I}_N\right)$, where $\mathbf{I}_N$ is the $\left(N\times N\right)$ identity matrix.

Denote $s_i = \left[s_i^{(x)},s_i^{(y)}\right]^T$ the dual-pol. symbol transmitted in the time step $i$, then an LMMSE estimate $\hat{s}_i= \left[\hat{s}_i^{(x)},\hat{s}_i^{(y)}\right]^T$  of $s_i$ can be obtained by the receiver as 
\begin{equation}\label{Eq10}
\hat{s}_{i}=\operatorname{E}\left\{s_{i}\right\}+\mathbf{w}_{i}^{H}\left[\mathbf{r}_{i}-\operatorname{E}\left\{\mathbf{r}_{i}\right\}\right],
\end{equation}

\noindent where $\mathbf{w}_{i}$ is an $\left(2N\times2\right)$ estimate of the Wiener filter coefficients
\begin{equation}\label{Eq11}
\mathbf{w}_{i}=\operatorname{cov}\left\{\mathbf{r}_{i}, \mathbf{r}_{i}\right\}^{-1} \operatorname{cov}\left\{\mathbf{r}_{i}, s_{i}\right\},
\end{equation}
\noindent where $\operatorname{cov}\{\mathbf{u}, \mathbf{v}\} \triangleq \operatorname{E}\left\{[\mathbf{u}-\operatorname{E}\{\mathbf{u}\}][\mathbf{v}-\operatorname{E}\{\mathbf{v}\}]^{H}\right\}$.

As discussed in \cite{Dejonghe2002}, in analogy with the classical SISO processing implemented using conventional \textit{a posteriori} probability (APP) algorithms based on \cite{Bahl1974}, the \textit{a priori} information about the symbols $s_i$ should not be included in the evaluation of its estimates $\hat{s}_i$. Hence, without \textit{a priori} information on $s_i$, we have $\operatorname{E}\{s_i\}= [0,0]^T$ and $\operatorname{E}\left\{\mathbf{r}_{i}\right\}=\mathbf{H}_i \overline{\mathbf{s}}_{i}$, and (\ref{Eq10}) simplifies to
\begin{equation}\label{Eq12}
\hat{s}_{i}=\mathbf{w}_{i}^{H}\left[\mathbf{r}_{i}-\mathbf{H}_i \overline{\mathbf{s}}_{i}\right],
\end{equation}

\noindent with $\overline{\mathbf{s}}_{i} = \left[\overline{\mathbf{s}}_{i}^{(x)},\overline{\mathbf{s}}_{i}^{(y)}\right]^T$, where
\begin{equation}\label{Eq13}
\overline{\mathbf{s}}_{i}^{(p)} \triangleq\left[\begin{array}{ccccccc}
\bar{s}_{i-N_{1}-L}^{(p)}, & \ldots, & \bar{s}_{i-1}^{(p)}, & 0, & \bar{s}_{i+1}^{(p)}, & \ldots, & \bar{s}_{i+N_{2}}^{(p)}
\end{array}\right],
\end{equation}
\noindent and $\overline{s}_{i}^{(p)} \triangleq \operatorname{E}\left\{s_{i}^{(p)}\right\}$.

Moreover, (\ref{Eq11}) can be written as 
\begin{equation}\label{Eq14}
\mathbf{w}_{i}=\left[\mathbf{H}_i \mathbf{R}_{\mathbf{s s}, i} \mathbf{H}_i^{H}+\sigma_{n}^{2} \mathbf{I}_{2N}\right]^{-1} \mathbf{h}_i\sigma_{s}^{2},
\end{equation}

\noindent where
\begin{equation}
\mathbf{h}_i \triangleq\left[\begin{array}{cc}
\mathbf{H}_i^{(xx)}\mathbf{e} & \mathbf{H}_i^{(xy)}\mathbf{e} \\
\mathbf{H}_i^{(yx)}\mathbf{e} & \mathbf{H}_i^{(yy)}\mathbf{e} \\
\end{array}\right]\nonumber,
\end{equation}

\noindent with $\mathbf{e}$ denoting a length-$(N + L)$ vector of all zeros, except for the $(N_1+L+1)th$ element, which is 1. Finally, under the independence assumption between the coded bits and between polarization components, the covariance matrix $\mathbf{R}_{\mathrm{ss}, i}$ is defined as $\mathbf{R}_{\mathrm{ss}, i} = \operatorname{diag}\left[\mathbf{R}_{\mathrm{ss}, i}^{(x)}, \mathbf{R}_{\mathrm{ss}, i}^{(y)}\right]$
where
\begin{equation}\label{Eq15}
\begin{array}{c}
\mathbf{R}_{\mathrm{ss}, i}^{(p)}=\left[\operatorname{var}\left\{s_{i-N_{1}-L}^{(p)}\right\}, \quad \ldots, \quad \operatorname{var}\left\{s_{i-1}^{(p)}\right\}, \quad \sigma_{s}^{2},\right. \\
\left.\operatorname{var}\left\{s_{i+1}^{(p)}\right\}, \quad \ldots, \quad \operatorname{var}\left\{s_{i+N_{2}}^{(p)}\right\}\right]
\end{array}.
\end{equation}

Note that, in the absence of prior information, the equalizer becomes the standard LMMSE equalizer
\begin{equation}\label{Eq16}
\mathbf{w}_{i}=\left[\sigma_{s}^{2}\mathbf{H}_i \mathbf{H}_i^{H}+\sigma_{n}^{2} \mathbf{I}_{2N}\right]^{-1} \mathbf{h}_i\sigma_{s}^{2},
\end{equation}
\begin{equation}\label{Eq17}
\hat{s}_{i}=\mathbf{w}_{i}^{H}\mathbf{r}_{i}.
\end{equation}
The complexity of the equalization is dominated by the inversion of the covariance matrix in (\ref{Eq14}) at each time step~$i$. In \cite{Tuchler2002}, a recursive procedure is presented that allows an efficient implementation leading to $O(N^2)$ complexity in the calculation of $\mathbf{w}_{i}$.
\\
\subsubsection{Extrinsic bit probability estimator}

In order to calculate the extrinsic soft information given the LMMSE estimates $\hat{s}_{i}$, first an equivalent AWGN channel assumption is made from the transmitter to the output of the equalizer, given by
\begin{equation}\label{Eq18}
\hat{s}_{i} = \mathbf{M}_i s_i + \eta_i,
\end{equation}

\noindent where $\mathbf{M}_i~=~\operatorname{diag}\left[\mu_{i}^{(x)}, \mu_{i}^{(y)}\right]$ is a scaling factor and the noise $\eta_{i}$ is distributed according to $\eta_{i}\sim~\mathcal{N}_{c}\left(0,\mathbf{N}_i \right)$, with $\mathbf{N}_i=\operatorname{diag}\left[\nu_{i}^{2(x)}, \nu_{i}^{2(y)}\right]$.

Following the derivations in \cite{Dejonghe2002}, the parameters $\mu_i$ and $\nu_i$ in (\ref{Eq18}) are calculated for each time step $i$ as a function of the equalizer coefficients as
\begin{align}
\mathbf{M}_i&\approx \mathbf{w}_{i}^{H} \mathbf{h},\label{Eq19}\\
\mathbf{N}_i&= \mathbf{M}_i \sigma_{s}^{2}-\mathbf{M}_i^{2} \sigma_{s}^{2}\label{Eq20},
\end{align}
\noindent with the off diagonal values of the matrix calculated in the right-hand side of (\ref{Eq19}) approximated to zero. From the equivalent channel model in (\ref{Eq18}), the following channel rule can be written for each polarization component $p$:
\begin{equation}\label{Eq21}
p\left(\hat{s}_{i}^{(p)} \mid s_{i}^{(p)}\right)=\frac{1}{\nu_{i}^{2(p)} \pi} \exp \left[-\frac{\left|\hat{s}_{i}^{(p)}-\mu_{i}^{(p)} s_{i}^{(p)}\right|^{2}}{\nu_{i}^{2(p)}}\right].
\end{equation}

The posterior symbol probability $P_{p}\left(s_{i}\right)$ is approximated by
\begin{equation}\label{Eq22}
P_{p}\left(s_{i}\right) \approx P\left(s_{i} \mid \hat{s}_{i}\right)=\frac{p\left(\hat{s}_{i} \mid s_{i}\right) P_{a}\left(s_{i}\right)}{p\left(\hat{s}_{i}\right)}.
\end{equation}

By definition, the extrinsic symbol probability $P_{e}\left(s_{i}\right)$ is proportional to ratio between posterior and prior symbol probabilities, and from (\ref{Eq22}) it follows that
\begin{equation}
P_{e}\left(s_{i}\right) \triangleq \kappa_{s} \frac{P_{p}\left(s_{i}\right)}{P_{a}\left(s_{i}\right)} \sim \frac{p\left(\hat{s}_{i} \mid s_{i}\right)}{p\left(\hat{s}_{i}\right)} \sim p\left(\hat{s}_{i} \mid s_{i}\right),
\end{equation}

\noindent where $\kappa_{s}$ is a normalization constant. From the extrinsic symbol probabilities, the extrinsic bit probabilities are obtained as 
\begin{equation}
P_{e}\left(b_{i}^{l}\right) \triangleq \kappa_{b} \frac{P_{p}\left(b_{i}^{l}\right)}{P_{a}\left(b_{i}^{l}\right)} \sim \sum_{s_{i}: b_{i}^{l}} P_{e}\left(s_{i}\right)\left[\prod_{r=1, \ldots, q \atop r \neq l} P_{a}\left(b_{i}^{r}\right)\right],
\end{equation}
\noindent where $\kappa_{b}$ is a normalization constant and the notation $s_{i}: b_{i}^{l}$ indicates the subset of all symbols in $\mathcal{S}$ with a given value of $b_{i}^{l}$. Thus, the probabilities $P_{e}\left(b_{i}^{l}\right)$ can be calculated using

\begin{equation}\label{Eq25}
P_{e}\left(b_{i}^{l}\right) \sim \sum_{s_{i}: b_{i}^{l}} p\left(\hat{s}_{i} \mid s_{i}\right)\left[\prod_{r=1, \ldots, q \atop r \neq l} P_{a}\left(b_{i}^{r}\right)\right].
\end{equation}

Not that only the \textit{a priori} probabilities of the other bits $b_{i}^{r} \; (r = 1,\dots,q; r\neq l)$ associated with the considered symbol $s_i$ are used to obtain the extrinsic probability of the bit $b_{i}^{l}$. That should allow for better demapping by avoiding possible mutual influence between encoded bits \cite{Dejonghe2002}.

Finally, using (\ref{Eq25}), the extrinsic bit L-values are calculated as
\begin{align}\label{Eq26}
L_e\left(b_i^l\right) &\approx\ln \frac{P_e\left(b_i^l=1\right)}{P_e\left(b_i^l=0\right)}\nonumber\\ 
 &= \ln \frac{\sum_{s_i: b_i^l=1} p\left(\hat{s}_i \mid s_i\right)\left[\prod_{r=1, \ldots, q \atop r \neq l} P_a\left(b_i^r\right)\right]}{\sum_{s_i: b_i^l=0} p\left(\hat{s}_i \mid s_i\right)\left[\prod_{r=1, \ldots, q \atop r \neq l} P_a\left(b_i^r\right)\right]}.
\end{align}

The SISO equalizer processing steps are summarized in the Algorithm~\ref{SISOalg}.

\begin{algorithm}
\SetAlgoLined
 \KwResult{Extrinsic bit L-values $L_e\left(b_i^l\right)$.}
 \KwIn{$\left\lbrace r_{i},\operatorname{E}\{s_i\},\operatorname{var}\{s_i\}, \mathbf{H}_i \right\rbrace_{i=1}^{m}$, prior bit L-values.} 
 \For{$i = 1;\ i \leq m;\ i = i + 1$}{
  get $\mathbf{r}_{i}$, $\overline{\mathbf{s}}_{i}$, $\mathbf{R}_{\mathbf{s s}, i}$, $\mathbf{H}_i$, $\mathbf{h}_i$\;
  calculate $\mathbf{w}_{i}$ (\ref{Eq14})\;
  calculate $\hat{s}_{i}$    (\ref{Eq12})\;  
  calculate $\mathbf{M}_i$ and $\mathbf{N}_i$ (\ref{Eq19}-\ref{Eq20})\;
  \For{$s_i \in \mathcal{S}$}{
      calculate $p(\hat{s}_{i}|s_{i})$ (\ref{Eq21})
  }
  \For{$l = 1;\ l \leq q;\ l = l + 1$}{
      calculate $L_e\left(b_i^l\right)$ (\ref{Eq26})
  }
 }
 \caption{SISO equalizer}
 \label{SISOalg}
\end{algorithm}
 \vspace{-0.5cm}
\subsection{Adaptive RLS channel estimator}\label{secII.B}

The adaptive channel estimation algorithm is implemented with the complex-valued 2$\times$2 adaptive filter described in (\ref{Eq27}),
\begin{equation}\label{Eq27}
 \begin{bmatrix}\hat{r}_{i}^{(x)} \\ \hat{r}_{i}^{(y)} \end{bmatrix} = \begin{bmatrix} \bm{h}_{i}^{(xx)H} & \bm{h}_{i}^{(xy)H} \\ \bm{h}_{i}^{(yx)H} & \bm{h}_{i}^{(yy)H}\end{bmatrix}\!\! \begin{bmatrix} \overline{\bm{s}}_{i}^{(x)} \\ \overline{\bm{s}}_{i}^{(y)}\end{bmatrix}.
\end{equation}

\noindent The RLS update of the coefficients \cite{Diniz2002} follows from (\ref{Eq28}) and (\ref{Eq29}), with
\begin{flalign} \label{Eq28}
\bm{\Sigma}_{i+1}^{(x)} &= \frac{1}{\lambda}\left[\bm{\Sigma}_i^{(x)} - \frac{\bm{\Sigma}_i^{(x)}\overline{\bm{s}}_{i}^{(x)}\overline{\bm{s}}_{i}^{(x)H}\bm{\Sigma}_i^{(x)}}{\lambda + \overline{\bm{s}}_{i}^{(x)H}\bm{\Sigma}_i^{(x)}\overline{\bm{s}}_{i}^{(x)}} \right], \nonumber\\
\bm{\Sigma}_{i+1}^{(y)} &= \frac{1}{\lambda}\left[\bm{\Sigma}_i^{(y)} - \frac{\bm{\Sigma}_i^{(y)}\overline{\bm{s}}_{i}^{(y)}\overline{\bm{s}}_{i}^{(y)H}\bm{\Sigma}_i^{(y)}}{\lambda + \overline{\bm{s}}_{i}^{(y)H}\bm{\Sigma}_i^{(y)}\overline{\bm{s}}_{i}^{(y)}} \right], 
\end{flalign}

\begin{flalign} \label{Eq29}
\bm{h}_{i+1}^{(xx)} &= \bm{h}_{i}^{(xx)} + e_{i}^{(x)*}\bm{\Sigma}_{i+1}^{(x)}\overline{\bm{s}}_{i}^{(x)},\nonumber\\
\bm{h}_{i+1}^{(xy)} &= \bm{h}_{i}^{(xy)} + e_{i}^{(x)*}\bm{\Sigma}_{i+1}^{(y)}\overline{\bm{s}}_{i}^{(y)},\nonumber\\
\bm{h}_{i+1}^{(yx)} &= \bm{h}_{i}^{(yx)} + e_{i}^{(y)*}\bm{\Sigma}_{i+1}^{(x)}\overline{\bm{s}}_{i}^{(x)},\nonumber\\
\bm{h}_{i+1}^{(yy)} &= \bm{h}_{i}^{(yy)} + e_{i}^{(y)*}\bm{\Sigma}_{i+1}^{(y)}\overline{\bm{s}}_{i}^{(y)},
\end{flalign}

\noindent where $(L+1)$ is the number of filter taps, $\overline{\bm{s}}_{i}^{(p)}~=~[\overline{s}_{i-d}^{(p)},...,\overline{s}_{i-d-L-1}^{(p)}]^T$, $p\in\{x,y\}$ and $d=\lfloor(L+1)/2\rfloor$ is the decision delay. The adaptive filter components are updated at every time step $i$ and they have the form $\bm{h}_i^{(k)}~=~\left[ h_{0}^{(k)}, h_{1}^{(k)},..., h_{L}^{(k)}\right]^T$, with $k\in \{xx,\; xy,\; yx,\; yy\}$. 

The $((L+1)\times (L+1))$ matrices $\bm{\Sigma}_i^{(x)}$ and $\bm{\Sigma}_i^{(y)}$ correspond to the inverse of the deterministic correlation matrix of the symbols in each polarization, and $\lambda$ is the forgetting factor. Finally, $e_{i}~=~\left[e_{i}^{(x)}, e_{i}^{(y)}\right]^T$ is the error between the estimated $\hat{r}_{i} = \left[\hat{r}_{i}^{(x)}, \hat{r}_{i}^{(y)}\right]^T$ and the actual channel output $r_{i} = \left[r_{i}^{(x)},r_{i}^{(y)}\right]^T$ at time step $i$.

The adaptive RLS channel estimation processing steps are summarized in the Algorithm~\ref{RLSChEst}.

\begin{algorithm}
\SetAlgoLined
 \KwResult{Estimated channel coefficients $\left\lbrace \mathbf{H}_i \right\rbrace_{i=1}^{m}$.}
 \KwIn{$\left\lbrace r_{i},\operatorname{E}\{s_i\} \right\rbrace_{i=1}^{m}$, $\lambda$.} 
 \For{$i = 1;\ i \leq m-1;\ i = i + 1$}{
  get $\mathbf{r}_{i}$, $\overline{\mathbf{s}}_{i}$, $\mathbf{H}_i$\;
  calculate $\hat{r}_{i}$ (\ref{Eq27})\;
  calculate $e_{i}~=~r_{i}-\hat{r}_{i}$\;
  update $\bm{\Sigma}_{i+1}^{(x)}$ and $\bm{\Sigma}_{i+1}^{(y)}$ (\ref{Eq28})\;  
  update and store $\bm{h}_{i+1}^{(k)}$, $k\in \left\lbrace xx, xy, yx, yy  \right\rbrace$  (\ref{Eq29})\; 
 }
 \caption{Adaptive RLS channel estimator}
 \label{RLSChEst}
\end{algorithm}
\vspace{-0.5cm}
\section{Numerical Simulations}\label{SecIII}

The receiver detailed in Sec.~\ref{SecII} is analyzed through Monte Carlo simulations. The numerical setup simulates a WDM system composed of eleven carriers modulated at $\mathrm{32~GBd}$ and disposed in a frequency grid with $\mathrm{37.5~GHz}$ of spacing. For every WDM carrier, the transmitted data bits are generated by encoding equiprobable random bit sequences with an LDPC code (rate $R=4/5$, block length $n=20480$) from the ARJ4A family \cite{Divsalar2005}. The coded bits in each LDPC frame are interleaved by a random permutation in order to avoid decoding issues due to possible error correlation introduced by the channel. The interleaver has the same length of the LDPC frame. After interleaving, the coded bits are Gray mapped to PM-64QAM/PM-256QAM symbols. Every Monte Carlo trial runs a sequence of symbols consisting of $\mathrm{18}$ LDPC code blocks per polarization component. Pilot-symbols for equalization and carrier phase recovery are uniformly inserted in between data symbols at a rate of $5\%$. The symbols are upsampled to $\mathrm{16~samples/symbol}$ and pulse shaped with a root-raised cosine (RRC) filter with a roll-off factor of $\mathrm{0.01}$. 

The fiber channel model assumes multiple $\mathrm{50~km}$ spans of SMF, with fiber attenuation compensated by Erbium-doped fiber amplifiers (EDFAs) with noise figure of $\mathrm{4.5~dB}$. The nonlinear WDM propagation is simulated applying the symmetric SSFM to solve the Manakov fiber model \cite{Marcuse1997}. The SSFM uses a fixed $\mathrm{100~m}$ step-size. The SMF fiber parameters of attenuation, nonlinear coefficient and chromatic dispersion are chosen to be $\mathrm{\alpha}=~0.2~\mathrm{dB/km}$, $\mathrm{\gamma}=~1.~3~\mathrm{W^{-1}km^{-1}}$, and $\mathrm{D}=~17~\mathrm{ps/nm/km}$, respectively. Local oscillator lasers are assumed to be ideal. Dynamic polarization effects, such as polarization mode dispersion, were not considered. 

At the receiver, the central channel is filtered out from the WDM signal by a band-pass filter (BPF). The signal undergoes downsampling to $\mathrm{2~samples/symbol}$, DBP (symmetric SSFM, $\mathrm{10~km}$ step-size, small enough to saturate the NLC performance) or electronic dispersion compensation (EDC), and RRC matched filtering (MF). The MF output is sent to a $\mathrm{T_s/2}$-fractionally spaced MIMO $\mathrm{2\times 2}$ normalized least mean squares (NLMS) adaptive equalizer. The filter lengths are fixed with $\mathrm{13~taps}$, since further increasing in the number of taps did not provide any reduction in the training MSE. The NLMS algorithm is set to operate in a pilot-based mode, i.e. the filter taps are updated only at the signaling intervals where pilot-symbols are located. The adaptive equalizer output is sent to the CPR, consisting of a decision directed phase-locked loop (DDPLL)\cite{Meyr1997}, which also takes advantage of the pilot-symbols to improve phase tracking. For more details about the DDPLL algorithm, see Sec.~III-C of \cite{Piels2015}. The CPR output is then sent to the turbo equalization processing. In the first iteration ($\mathrm{iter.~0}$), the SISO equalizer is bypassed. The bit L-values are calculated from the received symbols, de-interleaved, and sent directly to the LDPC decoder. Therefore, the performance metrics obtained after $\mathrm{iter.~0}$ define the reference receiver performance with DBP/EDC. Once the \textit{a posteriori} L-values are available at the output of the SISO decoder, the receiver starts its iterative processing, feeding the \textit{a posteriori} soft information back to the RLS channel estimator and to the SISO equalizer, as described in Sec.~\ref{SecII}. The pilot symbols are included in the iterative channel estimation and equalization steps by setting each pilot-mean and pilot-variance with the corresponding pilot-symbol and zero, respectively. The pilots are removed from the data after the LMMSE equalizer. Both RLS channel estimator filter and adaptive LMMSE filters are configured with $\mathrm{3~taps}$, i.e. $L=2$ and $N=3$ ($N_1=0,\; N_2=2$, see Sec.~\ref{SecII}), since any extra memory included in the processing did not provide significant improvement. For all tested configurations, setting the RLS forgetting factor $\mathrm{\lambda=0.99}$ provided the best performance. Before starting the RLS channel estimation, data-aided pre-convergence of the filter taps was performed using NLMS.

The figures of merit used to evaluate performance do not account for the symbols of the first three FEC code blocks, since those are used in the data-aided pre-convergence of the adaptive filters. Moreover, the last FEC frame is discarded to avoid  influence of filtering transients at the end of the trace. Therefore, for each Monte Carlo trial, the post-FEC BER is counted over $\mathrm{14}$ FEC code blocks. The same applies to SNR and AIR estimation. The LDPC decoding is performed using the standard belief propagation algorithm provided by MATLAB, which is based on the decoding algorithm presented by Gallager \cite{Gallagher1963}. The decoder is configured to perform at most $\mathrm{50}$ decoding iterations per frame, meaning that the decoding process is terminated either when all parity checks are satisfied or when the maximum number of iterations is reached. Here, the SNR always refers to the effectively received SNR estimated over a sequence of $m$ symbols, calculated at the output of the LMMSE equalizer as
\begin{equation}\label{Eq5}
\mathrm{SNR} \approx \frac{1}{2m}\sum_{p \in \{x,y\}}\sum_{i=1}^{m}\frac{|{s}_{i}^{(p)}|^2}{|\hat{s}_{i}^{(p)}-{s}_{i}^{(p)}|^2}.
\end{equation}

The information rates are estimated using generalized mutual information (GMI) \cite{Alvarado2018} under the assumption of an auxiliary additive complex-valued circular white Gaussian noise channel (\ref{Eq18}). Due to the feedback from the FEC decoder, the GMI can no longer be interpreted as an AIR. However, it is still useful in demonstrating the improvement of the quality of the L-values as a function of the number of iterations. The true meaning of the GMI and its relation to the AIR in the case of turbo equalization schemes is an interesting research matter out of this scope of this work. The motivation to present GMI curves is to provide a base for comparison with similar results already presented in the literature (see \ref{subSecI.a}). Predicting AIRs for turbo equalizers involves a careful analysis of extrinsic information transfer (EXIT) charts based on EXIT curves of the channel estimation/equalization step and the EXIT curve of the decoder. The tractability of this approach would depend on independence assumptions, e.g. between channel estimator/equalizer and FEC decoder operation, which may not hold for some schemes. To the best of our knowledge, for schemes such as the one proposed in this work, this is currently an open research problem on its own and the interested reader can find a useful discussion in \cite{Movahedian2015}.

In the following, the results of SNR versus transmission reach, and of GMI versus transmission reach are obtained for the optimal launch power of each receiver scheme.

\subsection{Performance of PM-256QAM with dispersion uncompensated transmission.}\label{secIII.A}
\begin{figure}[t]
\centering
\includegraphics[width=0.85\linewidth]{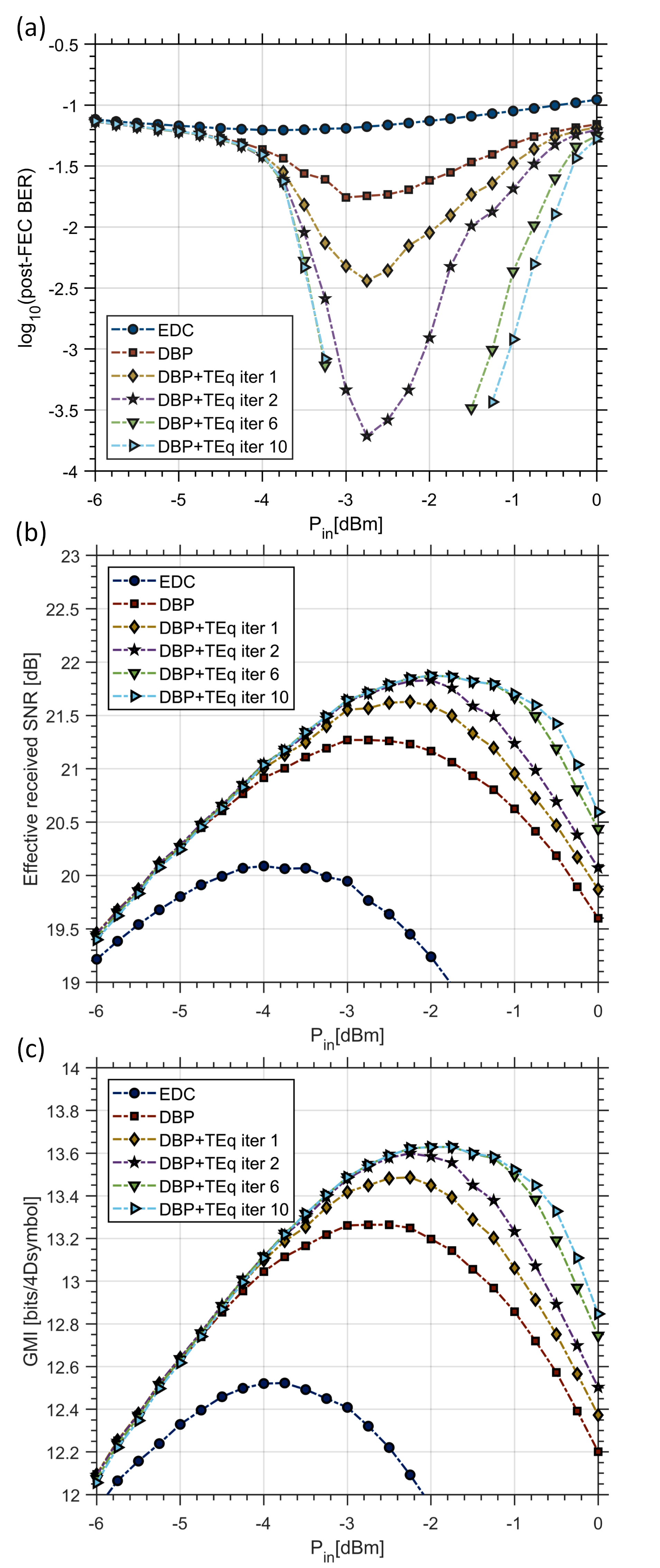}
\caption{Performance of the central channel versus fiber input power per WDM channel in a $\mathrm{11\times 32~GBd}$ PM-256QAM WDM system over $\mathrm{24\times 50~km}$ of dispersion uncompensated transmission with EDFA amplification. (a) Post-FEC BER; (b) Effectively received SNR; (c) Achievable rates estimated with GMI.}
\label{Fig2}
\vspace{-0.5cm}
\end{figure}

The performance curves shown in Fig.~\ref{Fig2} are obtained for the transmission distance fixed at $\mathrm{1200~km~(24\times 50~km)}$. In Fig.~\ref{Fig2}~(a), the evolution of the post-FEC BER as a function of the fiber input power per WDM channel is shown for different numbers of turbo iterations. The values displayed correspond to the average BER of $\mathrm{5}$ Monte Carlo runs, i.e. errors are counted over $\mathrm{70}$ LDPC code blocks.
 
Comparing the BER curves for EDC and DBP-only processing with the performance after turbo equalization, results clearly show that the proposed turbo equalizer allows the receiver to achieve lower post-FEC BER values. As the number of turbo iterations increases, better equalization is achieved and fewer bit errors are counted after the LDPC decoder. In this case, after $\mathrm{7}$ to $\mathrm{10}$ turbo iterations, the receiver was able to bring the BER down to $\mathrm{0}$. The SNR gain is higher at high power due to the relatively higher effect of the NLI, which is compensated by the turbo equalizer.

In Fig.~\ref{Fig2}~(b), the curves of the effectively received SNR versus the fiber input power per WDM channel are shown. The optimal launch power of $\mathrm{\approx-4~dBm}$ for EDC is shifted to $\mathrm{\approx-3~dBm}$ when DBP is applied. Including the turbo equalizer after DBP, the optimal launch power is further increased to $\mathrm{\approx-2~dBm}$. The SNR improvement provided by DBP over EDC is $\mathrm{\approx1.2~dB}$, while turbo equalization provides an extra $\mathrm{\approx0.6~dB}$. This improvement is notably higher than the previously reported  $\mathrm{0.2~dB}$ obtained with iterative schemes using the RLS adaptive filter as equalizer \cite{Golani2019, Golani2019_2, DaSilva2019}.

In Fig.~\ref{Fig2}~(c), the GMI curves versus fiber input power per WDM channel are presented. It can be seen that single-channel DBP increases the AIR by $\mathrm{\approx0.65~bits/4Dsymbol}$ respective to EDC, while the turbo equalizer allows an extra $\mathrm{\approx0.35~bits/4Dsymbol}$ on top of the DBP gain.

\begin{figure}[h]
\centering
\includegraphics[width=0.9\linewidth]{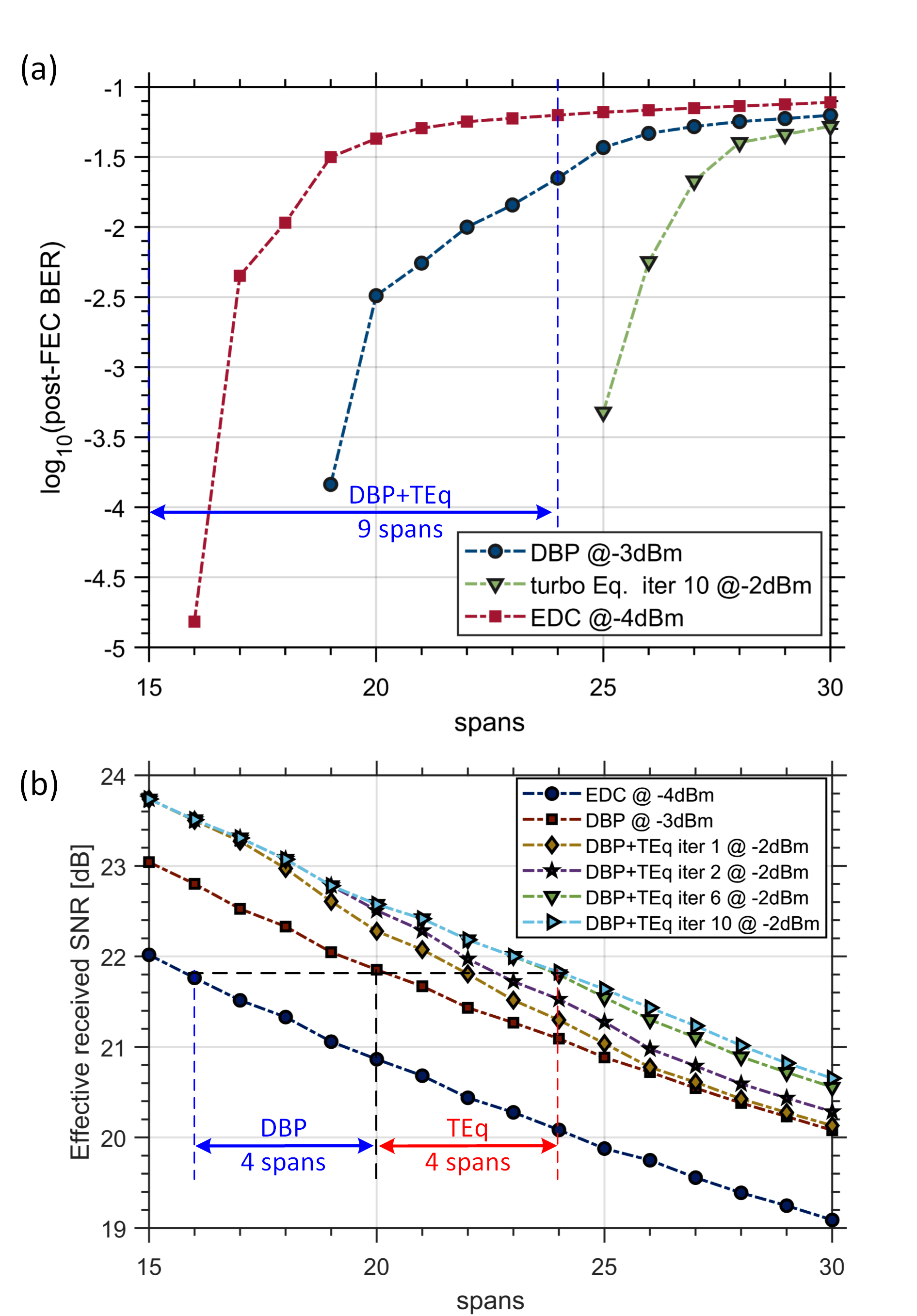}
\caption{Performance of the central channel in a $\mathrm{11\times 32~GBd}$ PM-256QAM WDM system transmitted over $\mathrm{50~km}$ spans at the optimal launch power for EDC, DBP, and DBP followed by turbo equalization. (a) Post-FEC BER versus transmission reach; (b) Effectively received SNR versus transmission reach.}
\label{Fig3}
\vspace{-0.2cm}
\end{figure}

In Fig.~\ref{Fig3}~(a), the post-FEC BER versus number of spans propagated by the WDM channels is shown. It is noted that with EDC no errors are counted up to span number $\mathrm{15}$, and with DBP up to span number $\mathrm{18}$. When turbo equalization is included after DBP, no errors are counted up to span number $\mathrm{24}$. Note that, in this high SNR regime, the power of time-varying inter-channel NLI dominates over the ASE. A fraction of the inter-channel NLI power comes in the form of nonlinear phase noise, breaking the Gaussian channel assumption used to calculate the L-values, which results in post-FEC BERs higher than what would be expected from a linear AWGN channel with the same SNR. Hence, even though it is clear that DBP provides the largest SNR gain, this gain does not translate directly in post-FEC BER improvement. Conversely, by using turbo equalization to compensate inter-channel NLI, the reach improvement is rather larger, because the channel model assumed by the decoder is better matched.

In Fig.~\ref{Fig3}~(b), the effectively received SNR evolution with the number of propagated spans is depicted. Note the transmission reach extension is doubled when the proposed turbo equalization scheme is included after DBP. The overall reach improvement obtained by DBP with turbo equalization with respect to EDC is $\approx 60\%$.

\begin{figure}[h]
\centering
\includegraphics[width=0.85\linewidth]{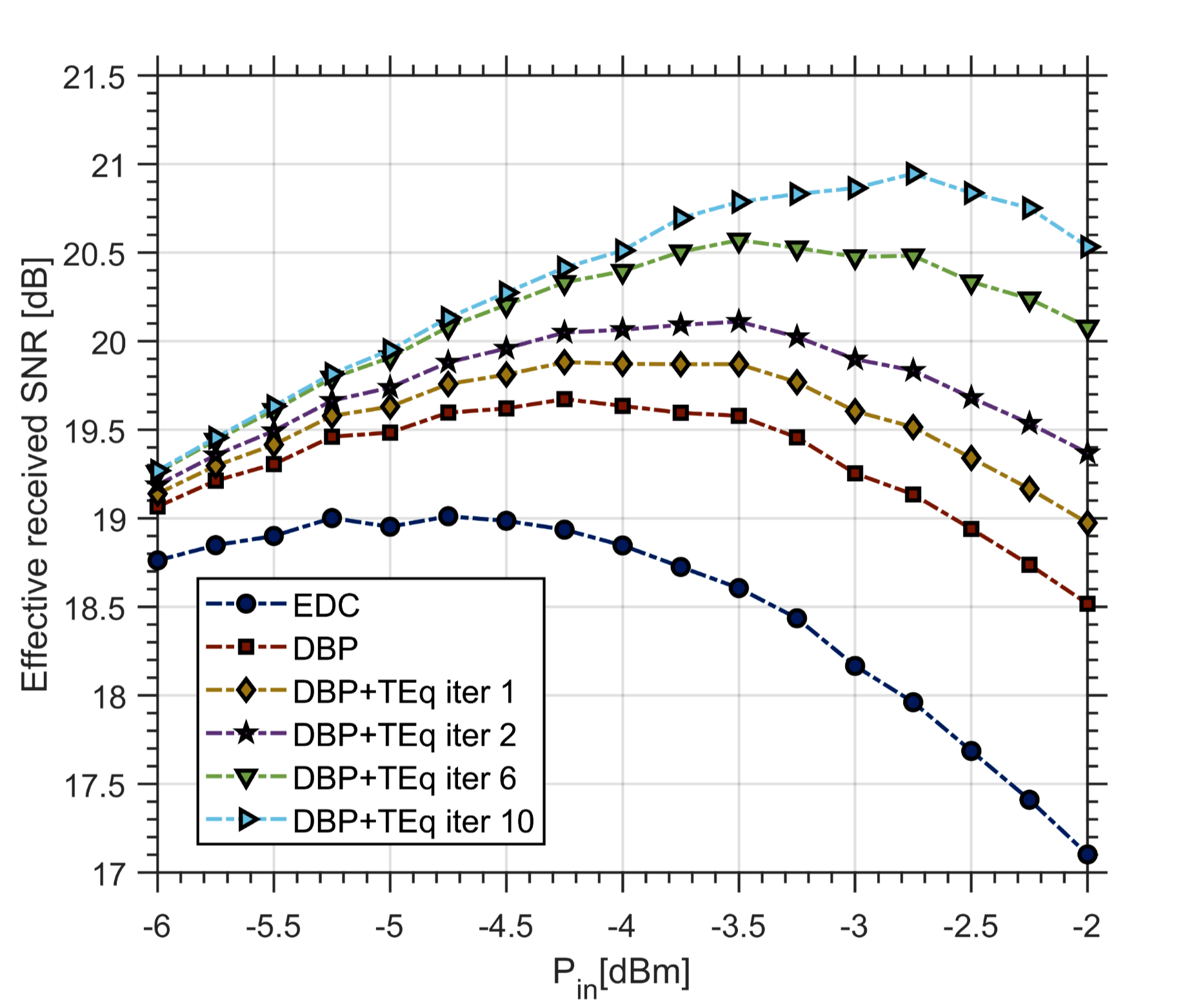}
\caption{Effectively received SNR of the central channel versus fiber input power obtained in the absence of adaptive equalization and CPR before the turbo equalizer for the same configuration described in Fig.~\ref{Fig2}~(b). } 
\label{Fig4}
\vspace{-0.2cm}
\end{figure}

To illustrate the impact of the standard DSP processing in the performance assessment of the turbo equalization, the same curves shown in Fig.~\ref{Fig2}~(b) are presented again in Fig.~\ref{Fig4}. 

The results in Fig.~\ref{Fig4} are obtained in the absence of adaptive equalization and CPR before the turbo equalizer. The first thing to notice is that removing those standard DSP blocks penalizes the performance with respect to Fig.~\ref{Fig2}~(b) by $\mathrm{\approx1.0~dB}$. Moreover, the gain provided by the turbo equalizer is now larger, around $\mathrm{1.3~dB}$, which agrees with the results presented in \cite{Golani2019}. That happens because a portion of the NLI compensated before by the adaptive equalizer and CPR is now left entirely to the turbo equalizer. Therefore, these results show that, without accounting for the standard DSP blocks at the receiver, the NLC performance of the turbo equalization may be overestimated.

\subsection{Performance of PM-64QAM with dispersion uncompensated transmission.}\label{secIII.B}

\begin{figure}[t]
\centering
\includegraphics[width=0.85\linewidth]{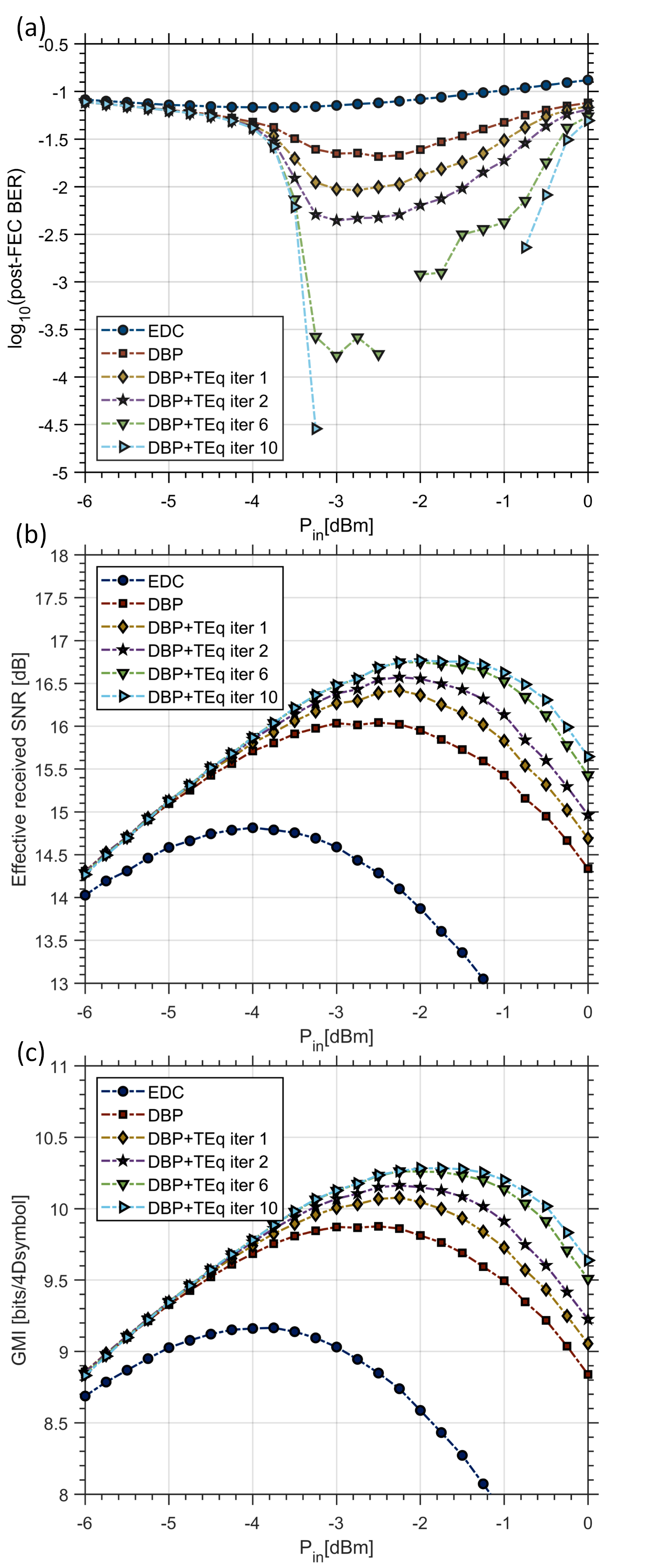}
\caption{Performance of the central channel versus fiber input power per WDM channel in a $\mathrm{11\times 32~GBd}$ PM-64QAM WDM system over $\mathrm{80\times 50~km}$ of dispersion uncompensated transmission with EDFA amplification. (a) Post-FEC BER; (b) Effectively received SNR; (c) Achievable rates estimated with GMI.}
\label{Fig5}
\vspace{-0.5cm}
\end{figure}

The curves shown in Fig.~\ref{Fig5} characterize the central channel performance of $\mathrm{11\times 32~GBd}$ PM-64QAM WDM system with the transmission distance fixed at $\mathrm{4000~km~(80\times 50~km)}$. Compared to the results presented in Fig.~\ref{Fig2}, similar improvements are observed, with slightly higher NLC gains from the turbo equalization, which is due to increased accumulation of NLI with a longer transmission distance. In this case, turbo equalization allows an SNR gain of $\mathrm{\approx0.7~dB}$, increasing GMI by $\mathrm{\approx0.4~bits/4Dsymbol}$. 

\begin{figure}[h]
\centering
\includegraphics[width=0.9\linewidth]{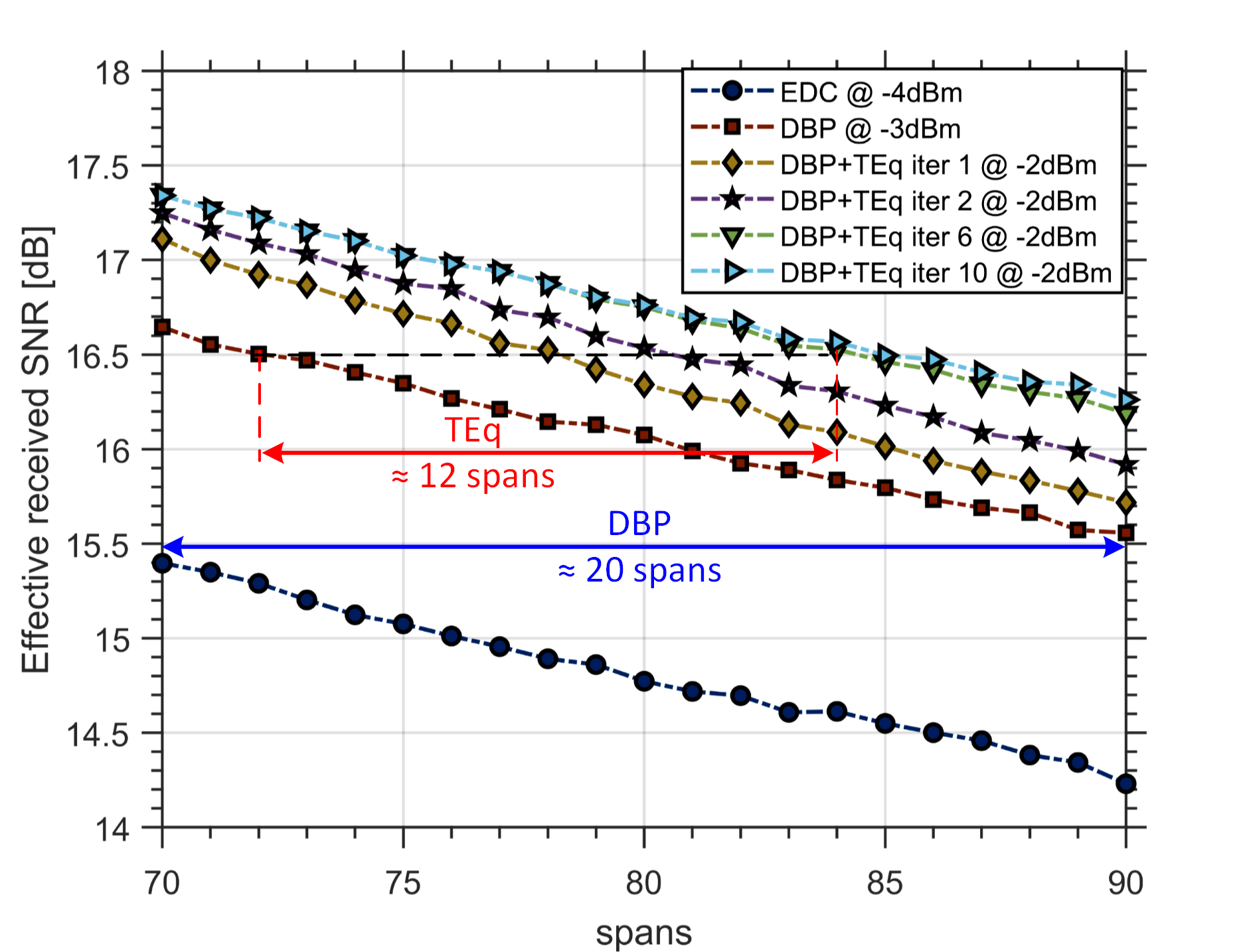}
\caption{Effectively received SNR versus transmission reach of the central channel in a $\mathrm{11\times 32~GBd}$ PM-64QAM WDM system transmitted over $\mathrm{50~km}$ spans at the optimal launch power for EDC, DBP, and DBP followed by turbo equalization.}
\label{Fig6}
\vspace{-0.5cm}
\end{figure}

Finally, as shown in Fig.~\ref{Fig6}, the reach improvement by using DBP ($\mathrm{\approx 20\; spans}$) with turbo equalization ($\mathrm{\approx 12\; spans}$) with respect to EDC is $\approx (20+12)/48 = 67\%$.
 
\section{Discussion and future work}\label{secIV}

As the authors in \cite{Golani2018,Golani2019} discuss, the RLS algorithm has reduced complexity relative to the Kalman algorithms used to perform adaptive channel estimation. A discussion about the similarities and differences of Kalman and RLS algorithms can be found in Chap. 17 of \cite{Diniz2002}. In \cite{Golani2018}, the complexity of the Kalman channel estimator is compared to the complexity of the RLS algorithm. The authors estimate that the computational complexity of the Kalman algorithm is higher than the RLS algorithm by a factor proportional to the cube of the memory length considered in the channel model. The Kalman algorithm proposed in \cite{Golani2018} is further extended in \cite{Golani2019}, where the channel estimation is carried out by an extended Kalman filter in combination with a Kalman smoother, which further increases the complexity of the channel estimation step. Since in this work we have only modified the function of RLS algorithm from adaptive channel equalizer to adaptive channel estimator, which only requires a swap on its input/output configuration, its computational complexity remains unaltered. Thus, it is expected that the complexity comparison presented in \cite{Golani2018,Golani2019} still holds.

The computational burden of the turbo equalization depends on the number of iterations between SISO equalizer and SISO decoder. In the simulations, it was observed that $\mathrm{\approx3}$ iterations were enough to achieve most of the improvement for a transmission reach around $\mathrm{1200~km}$, whereas twice as many iterations were required for distances around $\mathrm{4000~km}$. This behavior is intuitively explained considering that, with reduced effectively received SNR, one should expect performance degradation of the channel estimator and the LMMSE equalizer, which results in slower convergence of the iterative processing.

The FEC parameters were kept fixed in all scenarios investigated in this work. However, one should expect a trade-off between the SISO decoder performance and the overall improvement achieved by turbo equalization. Hence, it is yet to be investigated if and how the FEC specifications influence the NLC performance and if those parameters can be optimized to minimize the complexity of the turbo equalization. Moreover, as demonstrated in \cite{Yankov2021}, by using FEC codes capable of detecting block errors, it is possible to configure the receiver such that it triggers the iterative NLC only for blocks that could not be correctly decoded, thus reducing the complexity of the NLC. 

Although this has not been considered in this work, it has been demonstrated that probabilistic amplitude shaping (PAS) can have an impact on the effectively received SNR \cite{Fehenberger2020} due to the interplay of PAS block lengths and fiber nonlinear effects. The algorithms described in this paper are fully compatible with PAS, only requiring the probability mass function of the PAS symbols to be used as prior symbol distribution in the processing described in Algorithm~\ref{SISOalg}. Hence, the performance of the proposed scheme in PAS transceivers is to be investigated. Finally, performance bounds for such kind of iterative receivers for the nonlinear fiber channel are unknown, and it may be worth looking for ways to estimate the AIR limits for those receivers.
\vspace{-0.2cm}
\section{Conclusions}\label{secV}
An adaptive turbo equalization scheme for inter-channel NLC in polarization-multiplexed WDM systems was proposed and assessed via extensive numerical simulations. The processing at the receiver is carried by a SISO LMMSE equalizer and a low-complexity RLS adaptive channel estimator. The proposed scheme uses soft-information feedback from a SISO LDPC decoder as prior information to perform both channel estimation and equalization. Results were provided for $\mathrm{11\times 32~GBd}$ WDM system with 64 and 256QAM modulation carrying LDPC coded data. It is shown that turbo equalization is able to provide gains of up to $\mathrm{\approx0.7~dB}$ in SNR, and $\mathrm{\approx0.4~bits/4Dsymbol}$ in GMI. Transmission reach improvements of up to $\mathrm{\approx67\%}$ are observed by receivers combining single-channel DBP for intra-channel NLC and turbo equalization for inter-channel NLC. 

\vspace{-0.1cm}
\section*{Acknowledgements}
Edson P. da Silva is supported by the National Council for Scientific and Technological Development
(CNPq), Brazil, Grant Universal 432214/2018-6. Metodi P. Yankov is supported by the Danish National Research Foundation
(DNRF) (Centre of Excellence Silicon Photonics for Optical Communications (SPOC), ref.DNRF123).

\vspace{-0.1cm}

\bibliographystyle{IEEEtran}
\bibliography{Bibliography}

\begin{thebibliography}{10}
\providecommand{\url}[1]{#1}
\csname url@samestyle\endcsname
\providecommand{\newblock}{\relax}
\providecommand{\bibinfo}[2]{#2}
\providecommand{\BIBentrySTDinterwordspacing}{\spaceskip=0pt\relax}
\providecommand{\BIBentryALTinterwordstretchfactor}{4}
\providecommand{\BIBentryALTinterwordspacing}{\spaceskip=\fontdimen2\font plus
\BIBentryALTinterwordstretchfactor\fontdimen3\font minus
  \fontdimen4\font\relax}
\providecommand{\BIBforeignlanguage}[2]{{%
\expandafter\ifx\csname l@#1\endcsname\relax
\typeout{** WARNING: IEEEtran.bst: No hyphenation pattern has been}%
\typeout{** loaded for the language `#1'. Using the pattern for}%
\typeout{** the default language instead.}%
\else
\language=\csname l@#1\endcsname
\fi
#2}}
\providecommand{\BIBdecl}{\relax}
\BIBdecl
\renewcommand{\BIBentryALTinterwordstretchfactor}{4}

\bibitem{Dar2013}
R.~Dar \emph{et~al.}, ``{Properties of nonlinear noise in long,
  dispersion-uncompensated fiber links},'' \emph{Optics Express}, vol.~21,
  no.~22, p. 25685, 2013.

\bibitem{Poggiolini2014}
P.~Poggiolini \emph{et~al.}, ``{The GN-model of fiber non-linear propagation
  and its applications},'' \emph{Journal of Lightwave Technology}, vol.~32,
  no.~4, pp. 694--721, 2014.

\bibitem{Alvarado2018}
A.~Alvarado \emph{et~al.}, ``{Achievable Information Rates for Fiber Optics:
  Applications and Computations},'' \emph{Journal of Lightwave Technology},
  vol.~36, no.~2, pp. 424--439, 2018.

\bibitem{Essiambre2010}
R.~J. Essiambre \emph{et~al.}, ``{Capacity Limits of Optical Fiber Networks},''
  \emph{Journal of Lightwave Technology}, vol.~28, no.~4, pp. 662--701, 2010.

\bibitem{Cartledge2017}
J.~C. Cartledge \emph{et~al.}, ``{Digital signal processing for fiber
  nonlinearities [Invited]},'' \emph{Optics Express}, vol.~25, no.~3, p. 1916,
  2017.

\bibitem{Dar2015}
R.~Dar \emph{et~al.}, ``{Inter-channel nonlinear interference noise in WDM
  systems: Modeling and mitigation},'' \emph{Journal of Lightwave Technology},
  vol.~33, no.~5, pp. 1044--1053, 2015.

\bibitem{Ghazisaeidi2019}
A.~Ghazisaeidi, ``{Noise Analysis of Zero-Forcing Nonlinear Equalizers for
  Coherent WDM Systems},'' \emph{Journal of Lightwave Technology}, vol.~37,
  no.~6, pp. 1552--1559, 2019.

\bibitem{Roberts2006}
K.~Roberts \emph{et~al.}, ``{Electronic precompensation of optical
  nonlinearity},'' \emph{IEEE Photonics Technology Letters}, vol.~18, no.~2,
  pp. 403--405, 2006.

\bibitem{Ip2010a}
E.~Ip, ``{Nonlinear compensation using backpropagation for
  polarization-multiplexed transmission},'' \emph{Journal of Lightwave
  Technology}, vol.~28, no.~6, pp. 939--951, 2010.

\bibitem{Ip2010}
E.~M. Ip \emph{et~al.}, ``{Fiber Impairment Compensation Using Coherent
  Detection and Digital Signal Processing},'' \emph{Journal of Lightwave
  Technology}, vol.~28, no.~4, pp. 502--519, 2010.

\bibitem{Temprana2015a}
E.~Temprana \emph{et~al.}, ``{Overcoming Kerr-induced capacity limit in optical
  fiber transmission},'' \emph{Science}, vol. 348, no. 6242, pp. 1445--1448,
  jun 2015.

\bibitem{Secondini2013a}
M.~Secondini \emph{et~al.}, ``{Achievable information rate in nonlinear WDM
  fiber-optic systems with arbitrary modulation formats and dispersion maps},''
  \emph{Journal of Lightwave Technology}, vol.~31, no.~23, pp. 3839--3852,
  2013.

\bibitem{Secondini2014a}
------, ``{On XPM mitigation in WDM fiber-optic systems},'' \emph{IEEE
  Photonics Technology Letters}, vol.~26, no.~22, pp. 2252--2255, nov 2014.

\bibitem{Golani2018_2}
O.~Golani \emph{et~al.}, ``{Experimental characterization of nonlinear
  interference noise as a process of intersymbol interference},'' \emph{Opt.
  Lett.}, vol.~43, no.~5, pp. 1123--1126, Mar 2018.

\bibitem{Golani2018}
------, ``{Kalman-MLSE Equalization for NLIN Mitigation},'' \emph{Journal of
  Lightwave Technology}, vol.~36, no.~12, pp. 2541--2550, 2018.

\bibitem{Golani2019}
------, ``{NLIN Mitigation Using Turbo Equalization and an Extended Kalman
  Smoother},'' \emph{Journal of Lightwave Technology}, vol.~37, no.~9, pp.
  1885--1892, 2019.

\bibitem{DaSilva2019}
E.~P. {da Silva} \emph{et~al.}, ``{Perturbation-based FEC-Assisted iterative
  nonlinearity compensation for WDM systems},'' \emph{Journal of Lightwave
  Technology}, vol.~37, no.~3, pp. 875--881, 2019.

\bibitem{Paskov2019}
M.~{Paskov}, ``{Exploiting signal statistics in nonlinear digital signal
  proscessing},'' in \emph{Proc. of 45th European Conference on Optical
  Communication (ECOC 2019)}, 2019, pp. 1--4.

\bibitem{Arlunno2014}
V.~{Arlunno} \emph{et~al.}, ``{Turbo Equalization for Digital Coherent
  Receivers},'' \emph{Journal of Lightwave Technology}, vol.~32, no.~2, pp.
  275--284, 2014.

\bibitem{Pan2015}
C.~{Pan} \emph{et~al.}, ``{Optical Nonlinear Phase Noise Compensation for
  $9\times 32$ -Gbaud PolDM-16 QAM Transmission Using a Code-Aided
  Expectation-Maximization Algorithm},'' \emph{Journal of Lightwave
  Technology}, vol.~33, no.~17, pp. 3679--3686, 2015.

\bibitem{Golani2019_2}
O.~Golani \emph{et~al.}, ``{Equalization Methods for Out-of-Band Nonlinearity
  Mitigation in Fiber-Optic Communications},'' \emph{Applied Sciences}, vol.~9,
  no.~3, 2019.

\bibitem{KoikeAkino2020}
T.~{Koike-Akino} \emph{et~al.}, ``{Neural Turbo Equalization: Deep Learning for
  Fiber-Optic Nonlinearity Compensation},'' \emph{Journal of Lightwave
  Technology}, vol.~38, no.~11, pp. 3059--3066, 2020.

\bibitem{Tuchler2002}
M.~{Tuchler} \emph{et~al.}, ``{Minimum mean squared error equalization using a
  priori information},'' \emph{IEEE Transactions on Signal Processing},
  vol.~50, no.~3, pp. 673--683, 2002.

\bibitem{Yankov2015}
M.~P. {Yankov} \emph{et~al.}, ``{Low-Complexity Tracking of Laser and Nonlinear
  Phase Noise in WDM Optical Fiber Systems},'' \emph{Journal of Lightwave
  Technology}, vol.~33, no.~23, pp. 4975--4984, 2015.

\bibitem{Yankov2017}
------, ``{Nonlinear Phase Noise Compensation in Experimental WDM Systems With
  256QAM},'' \emph{Journal of Lightwave Technology}, vol.~35, no.~8, pp.
  1438--1443, 2017.

\bibitem{Fehenberger2015}
T.~{Fehenberger} \emph{et~al.}, ``{Compensation of XPM interference by blind
  tracking of the nonlinear phase in WDM systems with QAM input},'' in
  \emph{Proc. 2015 European Conference on Optical Communication (ECOC)}, 2015,
  pp. 1--3.

\bibitem{Dejonghe2002}
A.~{Dejonghe} \emph{et~al.}, ``{Turbo-equalization for multilevel modulation:
  an efficient low-complexity scheme},'' in \emph{Proc. 2002 IEEE International
  Conference on Communications. Conference Proceedings. ICC 2002 (Cat.
  No.02CH37333)}, vol.~3, 2002, pp. 1863--1867 vol.3.

\bibitem{Taubock2012b}
G.~Taub{\"{o}}ck, ``{Complex-valued random vectors and channels: Entropy,
  divergence, and capacity},'' \emph{IEEE Transactions on Information Theory},
  vol.~58, no.~5, pp. 2729--2744, 2012.

\bibitem{Alvarado2015}
A.~Alvarado \emph{et~al.}, ``{Replacing the Soft-Decision FEC Limit Paradigm in
  the Design of Optical Communication Systems},'' \emph{Journal of Lightwave
  Technology}, vol.~33, no.~20, pp. 4338--4352, oct 2015.

\bibitem{Bahl1974}
L.~{Bahl} \emph{et~al.}, ``{Optimal decoding of linear codes for minimizing
  symbol error rate (Corresp.)},'' \emph{IEEE Transactions on Information
  Theory}, vol.~20, no.~2, pp. 284--287, 1974.

\bibitem{Diniz2002}
P.~S. Diniz, \emph{{Adaptive Filtering: Algorithms and Practical Implementation
  (2nd ed.)}}.\hskip 1em plus 0.5em minus 0.4em\relax Springer, 2002.

\bibitem{Divsalar2005}
D.~{Divsalar} \emph{et~al.}, ``{Protograph based LDPC codes with minimum
  distance linearly growing with block size},'' in \emph{Proc. GLOBECOM '05.
  IEEE Global Telecommunications Conference, 2005.}, vol.~3, 2005, pp. 5 pp.--.

\bibitem{Marcuse1997}
D.~{Marcuse} \emph{et~al.}, ``{Application of the Manakov-PMD equation to
  studies of signal propagation in optical fibers with randomly varying
  birefringence},'' \emph{Journal of Lightwave Technology}, vol.~15, no.~9, pp.
  1735--1746, 1997.

\bibitem{Meyr1997}
H.~Meyr \emph{et~al.}, \emph{\BIBforeignlanguage{eng}{{Digital communication
  receivers, volume 2: synchronization, channel estimation and signal
  processing}}}.\hskip 1em plus 0.5em minus 0.4em\relax Wiley, 1997.

\bibitem{Piels2015}
M.~Piels \emph{et~al.}, ``{Laser Rate Equation-Based Filtering for Carrier
  Recovery in Characterization and Communication},'' \emph{Journal of Lightwave
  Technology}, vol.~33, no.~15, pp. 3271--3279, 2015.

\bibitem{Gallagher1963}
R.~G. Gallagher, \emph{{Low-density parity-check codes}}.\hskip 1em plus 0.5em
  minus 0.4em\relax MIT Press, 1963.

\bibitem{Movahedian2015}
A.~Movahedian \emph{et~al.}, ``{On the capacity of iteratively estimated
  channels using LMMSE estimators},'' \emph{IEEE Transactions on Vehicular
  Technology}, vol.~64, no.~1, pp. 97--107, 2015.

\bibitem{Yankov2021}
M.~P. {Yankov} \emph{et~al.}, ``{Block Error Detection Driven Nonlinearity
  Compensation for Optical Fiber Communications},'' \emph{IEEE Photonics
  Technology Letters}, vol.~33, no.~9, pp. 461--464, 2021.

\bibitem{Fehenberger2020}
T.~Fehenberger \emph{et~al.}, ``{Analysis of Nonlinear Fiber Interactions for
  Finite-Length Constant-Composition Sequences},'' \emph{J. Lightwave
  Technol.}, vol.~38, no.~2, pp. 457--465, Jan 2020.

\end{thebibliography}

%
%
\vfill


\end{document}